\documentclass[twocolumn,prb,showpacs,preprintnumbers,amsmath,amssymb]{revtex4}
\usepackage{graphicx}% Include figure files
\usepackage{dcolumn}% Align table columns on decimal point
\usepackage{bm}% bold math
\usepackage{epsfig}
\begin{document}

%\preprint{APS/123-QED}

\title{An Eshelby model for the highly viscous flow}% Force line breaks with \\

\author{U. Buchenau}
\email{buchenau-juelich@t-online.de}
\affiliation{%
Institut f\"ur Festk\"orperforschung, Forschungszentrum J\"ulich\\
Postfach 1913, D--52425 J\"ulich, Federal Republic of Germany}%
\date{September 24, 2007}% It is always \today, today,

\begin{abstract}
The shear flow and the dielectric $\alpha$-process in molecular glass formers is modeled in terms of local structural rearrangements which reverse a strong local shear. Using Eshelby's solution of the corresponding elasticity theory problem (J. D. Eshelby, Proc. Roy. Soc. {\bf A241}, 376 (1957)), one can calculate the recoverable compliance and estimate the lifetime of the symmetric double-well potential characterizing such a structural rearrangement. A full modeling of the shear relaxation spectra requires an additional parametrization of the barrier density of these structural rearrangements. The dielectric relaxation spectrum can be described as a folding of these relaxations with the Debye process.
\end{abstract}


\maketitle

\section{Introduction}

From an experimental point of view, broadband dielectric spectroscopy \cite{loidl} is the most versatile method to study the flow process in undercooled molecular liquids. However, the relation between dielectric relaxation and shear flow is not yet clear. The classical Debye picture and its extension to viscoelasticity \cite{gemant,dimarzio} considers the molecule as a small sphere immersed in the viscoelastic liquid. It predicts a slow dielectric decay, about a factor of ten slower than the one found in experiment \cite{boettcher,chang}. A thorough quantitative analysis of dielectric and shear data in seven glass formers \cite{niss} showed a general qualitative agreement with the extended Debye scheme, but a rather poor quantitative fit.

One cannot help feeling that the extended Debye scheme mistreats the structural rearrangements of the highly viscous fluid. The Debye relaxation time of the molecular orientation is usually longer than the Maxwell relaxation time of the shear stress, the more so the larger the molecular volume is.  No such retardation is expected for a local structural rearrangement, which ought to be characterized by the same relaxation time for shear and dielectrics. Thus the modeling should rather separate the viscous effects from those of the structural rearrangements, both in shear and dielectrics. Also, in the present unsettled state of understanding of the highly viscous flow, with many different ideas and recipes \cite{ngai,dyre,gotz,granato,avramov}, an attempt to understand the viscous flow itself in terms of a sequence of structural rearrangements in time seems legitimate. This is the purpose of the present paper.

In order to contribute effectively to the flow, the structural rearrangement should change the shape of the rearranged region in the direction of the flow. This implies a strained state of the embedding matrix, against the flow direction before the jump and in flow direction after. This mechanism will be explained in detail in the next section, section II. It leads to a finite lifetime of the corresponding double-well potential, because the surrounding matrix is itself able to flow. This lifetime implies a specific cutoff function for the barrier density of the structural rearrangements. The comparison with experiment in section III requires a specification of the barrier density, which does not follow immediately from the picture. As we will see, this requires three parameters even if there is no Johari-Goldstein secondary relaxation peak, and three more if such a peak is present. Such a large number of parameters makes it difficult to check the validity of the model with any certainty from the rather broad and featureless relaxation spectra.

\section{Eshelby model for the highly viscous flow}

\subsection{Shear strain defects}

%%%%%%%%%%%%%%%%%%%%% begin figure %%%%%%%%%%%%%%%%%%%%%%%%%%%%%%%%%%%%%
\begin{figure}[b]
\hspace{-0cm} \vspace{0cm} \epsfig{file=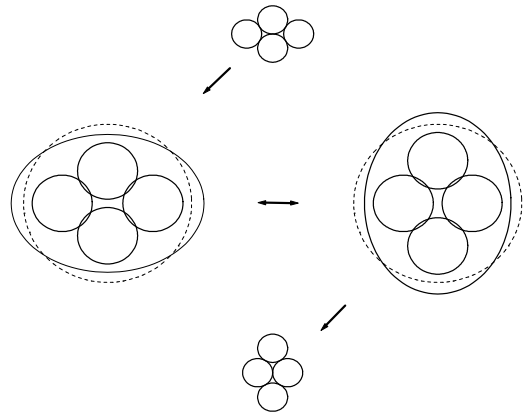,width=8cm,angle=0} \vspace{0cm}\caption{Elementary flow process (schematic), involving a rearrangement of four closely packed spherical atoms or molecules. The dashed lines show the shape of the hole containing the four molecules in the unstrained matrix, the continuous lines the shape of the hole in the strained matrix.}
\end{figure}
%%%%%%%%%%%%%%%%%%%%% end figure %%%%%%%%%%%%%%%%%%%%%%%%%%%%%%%%%%%%%%%

The central concept of this model is a structural rearrangement of a limited region in the sample which changes its shape to a sheared one. A simple example of such a rearrangement is the special case of four closely packed spherical atoms or molecules shown in Fig. 1. 

Imagine that the flow occurs by expansion in the vertical direction and a contraction of the horizontal direction of the figure. After a while, the four molecules at the top of the figure will no longer experience an adapted matrix, but rather a matrix which tends to compress them in the horizontal direction and to tear them apart in the vertical one. Since the four molecules have the alternative of the equivalent close packing shown at the bottom of the figure, there is a moment in time at which one has a symmetric double-well potential for molecules and matrix, the situation of the center of the figure.

The Eshelby model described here supposes that the elementary flow process in molecular liquids is of this nature, not necessarily restricted to four molecules (in fact, the quantitative discussion of section IV will show that the elementary flow process must be more complicated), but involving a structural rearrangement by a thermally activated jump between two stable locally ordered configurations with opposite shear strain with respect to the embedding matrix.

The physical problem of a small piece of matter able to transform to a sheared shape within an elastic matrix has been treated fifty years ago by several authors, notably by J. D. Eshelby \cite{eshelby,mura}.

Here, we translate Eshelby's result into the usual convention, in which the shear angle $e$ and the shear stress $\sigma$ are related by $\sigma=G e$ ($G$ infinite frequency shear modulus) at short times and by $\sigma=\eta \dot{e}$ ($\eta$ viscosity) in the long time limit. Let $v$ be the volume of the spherical inclusion and $e_i$ the shear angle difference between its two stable configurations (in the example of Fig. 1, $e_i=\pi/3$). Then the energy of the two equally strained configurations of the center of Fig. 1 is
\begin{equation}\label{ea}
	E_{a}=\frac{\gamma}{8} Gve_i^2.
\end{equation}
The coefficient $\gamma$ is given by
\begin{equation}\label{gamma}
\gamma=\frac{7-5\sigma_P}{15(1-\sigma_P)},	
\end{equation}
where $\sigma_P$ is Poisson's ratio. Since Poisson's ratio lies between 0.1 and 0.33 for the known glasses, $\gamma$ lies between 0.48 and 0.533, close to 1/2.

Eshelby's solution divides the energy into two almost equal parts, one located in the inclusion and one outside. Their ratio is $\gamma/(1-\gamma)$. The inclusion would have to distort by $e_i/2$ to fit exactly into the unstrained hole. In the four-atom case of Fig. 1, this is the saddle point energy for the thermally activated jump between the two stable configurations. It is considerably less than twice the energy of the two stable Eshelby minima because the saddle point has a lower energy than the harmonic extrapolation. Thus, in our simple four-atom example, the barrier height $V$ is lower than the formation energy $E_a$ of the symmetric double-well potential. We will see later that this cannot be true for the defects which actually destabilize the amorphous solid.

The Eshelby treatment supplies as well the interaction energy of the strain defect with the external stress component $\sigma$ oriented along the shear strain $e_i$
\begin{equation}\label{eint}
	E_{int}=-\frac{\sigma ve_i}{2}.
\end{equation}
This implies that the asymmetry $\Delta$ between the two minima changes by $\sigma ve_i/2$ in the presence of an external stress $\sigma$, or by $eGve_i/2$ in the presence of an applied external strain $e$.

\subsection{Double-well potential lifetime and cutoff function}

The lifetime $\tau_c$ of such a strain defect in the viscous liquid is estimated in a simple mean-field consideration. The shear stress outside, where other and independent strain defects exist, can decay according to the Maxwell relaxation time
\begin{equation}\label{maxwell}
	\tau_m=\frac{\eta}{G}.
\end{equation}
We assume that the energy inside the inclusion can only decay by the yielding of the surrounding matrix. Then the energy which has to decay is about twice the outside energy, which implies that $\tau_c$ is about twice $\tau_m$. More accurately, one has to take into account that the outside energy is not pure shear; about ten percent of it is compressional energy, which cannot be expected to decay. Thus one should have
\begin{equation}\label{tauc}
\tau_c=2.2\tau_m.
\end{equation}

The double-well potential lifetime $\tau_c$ determines the cutoff-barrier $V_c$ of the strain defects participating in the flow process according to the Arrhenius equation
\begin{equation}\label{tauv}
	\tau_V=\tau_0\exp(V/k_BT),
\end{equation}
where $T$ is the temperature and $\tau_0$ is a microscopic lifetime of $10^{-13}$ s.

At the critical energy barrier $V_c$, the potential decay begins to become faster than the jumps over the barrier. If the barrier is higher than $V_c$, the double-well potential begins to flow away before the population of the two minima can equilibrate. Therefore these higher barriers will not participate in the flow process.

Since one has two competing equilibration processes, the potential decay with the time constant $\tau_c$ and the thermally activated jumps over the barrier with the time constant $\tau_V$, the contribution of the jump mechanism is given by the cutoff function $l_c(V)$
\begin{equation}\label{lcv}
l_c(V)=\frac{\tau_c}{\tau_c+\tau_V}=\frac{1}{1+\exp((V-V_c)/k_BT)}.
\end{equation}

\subsection{Stationary flow and recoverable compliance}

In a stationary flow $\dot{e}$, the double-well potential asymmetry $\Delta$ changes continuously. If we ascribe $\dot{e}$ to the continuous passage of inclusions from one stable configuration to the other, it must result from an integral over all these processes. The constant flow $\dot{e}$ can be determined by counting all strain defects passing through the asymmetry zero
\begin{equation}
	\dot{e}=\dot{e}\int_0^\infty dV\frac{G(ve_i)^2}{4}\frac{n(V,0)}{5}l_c(V),
\end{equation}
The factor $1/5$ results from the averaging over the different strain orientations and $l_c(V)$ is the cutoff at the barrier $V_c$ from eq. (\ref{lcv}). Thus
\begin{equation}\label{nvint}
\int_0^\infty dV\frac{(ve_i)^2n(V,0)}{20}l_c(V)=\frac{1}{G}.
\end{equation}

Equation (\ref{nvint}) can be used to calculate the contribution of the shear strain defects to the shear compliance. To do this, consider first the free energy $F=-k_BT\ln Z$ of a single strain defect with asymmetry $\Delta$. The partition function $Z$ is
\begin{equation}
Z=2\cosh{\frac{\Delta}{2k_BT}}.
\end{equation}
Since the asymmetry $\Delta$ changes by $\sigma ve_i/2$ if one applies a stress $\sigma$ in the direction of $e_i$, one has a contribution to the shear compliance determined by
\begin{equation}\label{single}
\frac{\partial^2 F}{\partial\sigma^2}=-\frac{v^2e_i^2}{16k_BT\cosh^2{\Delta/2k_BT}}.
\end{equation}

In order to get the full recoverable compliance $J_e^0$ (the compliance after subtraction of the viscous flow contribution \cite{ferry}), one has to integrate over the barrier heights $V$ and the asymmetries $\Delta$. Here, we assume that the dependence of $n(V,\Delta)$ on $\Delta$ is given by the Boltzmann factor $\exp(-F/k_BT)$
\begin{equation}
n(V,\Delta)=n(V,0)\cosh{\frac{\Delta}{2k_BT}}.
\end{equation}

The equation for the recoverable compliance is
\begin{equation}
J_e^0=\frac{1}{G}+\int_0^\infty\int_{-\infty}^\infty dVd\Delta\frac{v^2e_i^2n(V,0)}{80k_BT\cosh{\Delta/2k_BT}}.
\end{equation}
The asymmetry can be integrated out. With equation (\ref{nvint}), one finds for the relaxational part of the recoverable compliance
\begin{equation}\label{j0e}
GJ_e^0-1\equiv f_0=\frac{\pi}{2}.
\end{equation}
The comparison to experiment in section III shows larger values. This will be discussed in section IV. 

\section{Comparison to experiment}

\subsection{The parametrization of the barrier density}

%%%%%%%%%%%%%%%%%%%%% begin figure %%%%%%%%%%%%%%%%%%%%%%%%%%%%%%%%%%%%%
\begin{figure}[b]
\hspace{-0cm} \vspace{0cm} \epsfig{file=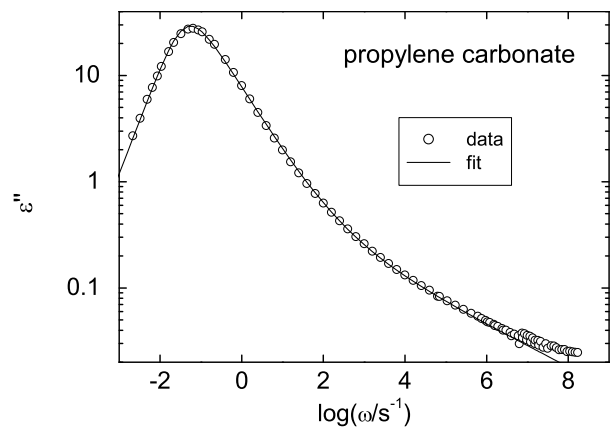,width=8cm,angle=0} \vspace{0cm} \caption{The dielectric data \cite{pc} of propylene carbonate at 158 K to which the barrier density was adapted.}
\end{figure}
%%%%%%%%%%%%%%%%%%%%% end figure %%%%%%%%%%%%%%%%%%%%%%%%%%%%%%%%%%%%%%%

The comparison to experiment requires the knowledge of the barrier density $n(V,0)$, which is difficult to model, particularly for glassforming systems, which need to be complex to avoid crystallization. Therefore we describe $n(V,0)$ or, more accurately, its product with the average square of the coupling constant, by the pragmatic form
\begin{equation}\label{lv}
\frac{l(V)}{l_nl_c(V)}=\exp{\frac{\beta(V-V_c)}{k_BT}}+\frac{1}{24}\exp{\frac{(V-V_c)}{5k_BT}}.
\end{equation}
The normalization factor $l_n$ is determined by the normalization condition
\begin{equation}\label{norm}
\int_0^\infty dVl(V)=1.
\end{equation}
The parameter $\beta$ has a close correspondence to the Kohlrausch-$\beta$ of approximately 0.5 of the Kohlrausch form $\exp{-t^\beta}$, the most popular fitting form for glassy relaxation. It supplies the slope $\omega^{-\beta}$ at frequencies shortly above the $\alpha$-peak frequency. The second term in eq. (\ref{lv}) provides a small curvature of this slope toward very high frequencies; we will see that one needs it. The prefactor of this term and the exponent are in fact additional parameters, which were fitted to the dielectric data of propylene carbonate \cite{pc} in Fig. 2. But the form is purely pragmatic, without any physical significance, with the exception of the cutoff. Note that this cutoff is not very sharp; therefore the $\beta$-values one obtains tend to be larger than those of a Kohlrausch fit, between 0.5 and 0.7.

If the glass former exhibits a pronounced secondary peak (Johari-Goldstein peak) one needs to add a gaussian with three parameters to the form of eq. (\ref{lv}). In this case, one has to reckon with different peak amplitudes in shear and dielectrics \cite{bu2007}.

The complex shear compliance, from which the complex modulus $G(\omega)$ can be easily calculated by inversion, is given by
\begin{equation}\label{jp}
GJ'(\omega)=1+(GJ_e^0-1)\int_0^\infty l(V)\frac{1}{1+\omega^2\tau_V^2} dV
\end{equation}
and
\begin{equation}\label{jpp}
GJ''(\omega)=(GJ_e^0-1)\int_0^\infty l(V)\frac{\omega\tau_V}{1+\omega^2\tau_V^2} dV+\frac{1}{\omega\tau_m}.
\end{equation}
Note that here appears the factor $f_0=GJ_e^0-1$, which according to eq. (\ref{j0e}) should be $\pi/2$. $\tau_m$ is the Maxwell time $\eta/G$.

If one deals with a single type of double-well potentials, they should show up with the same $l(V)$ in the shear compliance and in the dielectric susceptibility. The only difference lies in the viscous effects, which in the dielectric case should lead to a relaxation with the Debye relaxation time $\tau_D$
\begin{equation}\label{taud}
\tau_D=\frac{4\pi\eta r_H^3}{k_BT},
\end{equation}
where $r_H$ is the hydrodynamic radius of the molecule.

Since structural rearrangements and Debye relaxation must be considered as independent processes influencing the same molecule, one has to fold the two processes in frequency or to multiply them in time. Thus one can use the same $l(V)$ as in the shear case, replacing the $\tau_V$ of the Arrhenius equation, eq. (\ref{tauv}), by the shorter relaxation time $\tau_v$
\begin{equation}
\tau_v=\frac{\tau_V\tau_D}{\tau_V+\tau_D}.
\end{equation}

With this definition, the dielectric susceptibility (after subtraction of the conductivity contribution) reads  
\begin{equation}\label{ep}
\frac{\epsilon'(\omega)-\epsilon_\infty}{\epsilon(0)-\epsilon_\infty}=\int_0^\infty l(V)\frac{1}{1+\omega^2\tau_v^2} dV
\end{equation}
and
\begin{equation}\label{epp}
\frac{\epsilon''(\omega)}{\epsilon(0)-\epsilon_\infty}=\int_0^\infty l(V)\frac{\omega\tau_v}{1+\omega^2\tau_v^2} dV.
\end{equation}
Here $\epsilon(0)$ is the static dielectric susceptibility, $\epsilon_\infty$ is the real part of $\epsilon(\omega)$ in the GHz range (larger than $n^2$, the square of the refractive index, because of vibrational contributions \cite{bzow}).
  
%%%%%%%%%%%%%%%%%%%%% begin figure %%%%%%%%%%%%%%%%%%%%%%%%%%%%%%%%%%%%%
\begin{figure}[b]
\hspace{-0cm} \vspace{0cm} \epsfig{file=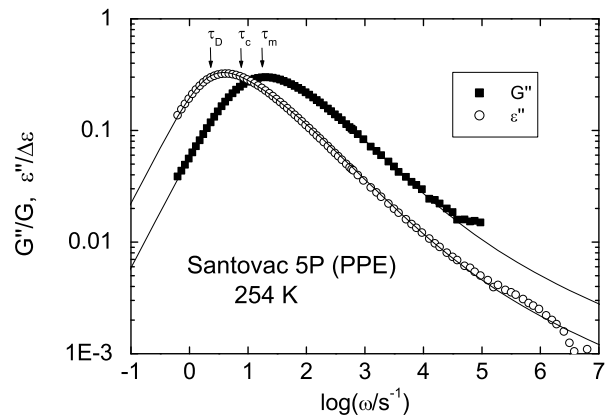,width=8cm,angle=0} \vspace{0cm} \caption{Model fits (continuous lines) of $G''(\omega)$ and $\epsilon''(\omega)$ (Roskilde data \cite{niss}) for a diffusion pump oil at 254 K.}
\end{figure}
%%%%%%%%%%%%%%%%%%%%% end figure %%%%%%%%%%%%%%%%%%%%%%%%%%%%%%%%%%%%%%%

These equations for the real and imaginary part of the dielectric constant contain one implicit assumption which might be wrong: That the same double-well potentials which allow the sample to flow are able to fully equilibrate the molecular orientation. We will come back to this point in the discussion.

\subsection{Examples}

The first example is Santovac 5P or PPE (polyphenylene), a diffusion pump oil, consisting of a short chain of five phenyl rings connected by oxygens. The measurements were supplied by the Roskilde group \cite{niss}.

%%%%%%%%%%%%%%%%%%%%% begin figure %%%%%%%%%%%%%%%%%%%%%%%%%%%%%%%%%%%%%
\begin{figure}[b]
\hspace{-0cm} \vspace{0cm} \epsfig{file=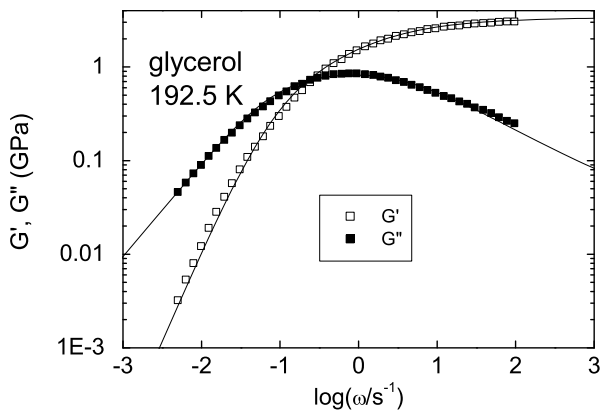,width=8cm,angle=0} \vspace{0cm} \caption{Fits of the shear data of Schr\"oter and Donth \cite{donthg} of glycerol.}
\end{figure}
%%%%%%%%%%%%%%%%%%%%% end figure %%%%%%%%%%%%%%%%%%%%%%%%%%%%%%%%%%%%%%%

%%%%%%%%%%%%%%%%%%%%% begin figure %%%%%%%%%%%%%%%%%%%%%%%%%%%%%%%%%%%%%
\begin{figure}[b]
\hspace{-0cm} \vspace{0cm} \epsfig{file=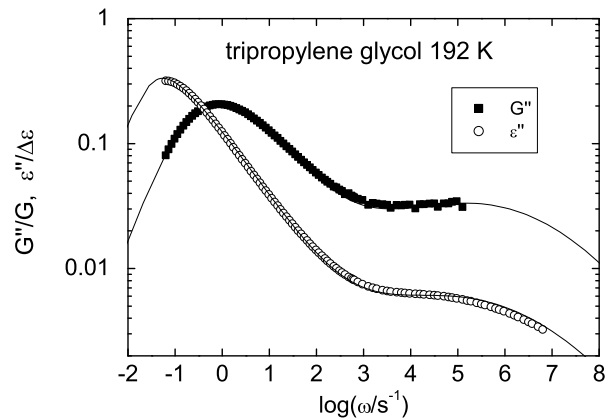,width=8cm,angle=0} \vspace{0cm} \caption{Fits of $G''(\omega)$ and $\epsilon''(\omega)$ (Roskilde data \cite{niss}) for tripropylene glycol, an example with a pronounced Johari-Goldstein peak.}
\end{figure}
%%%%%%%%%%%%%%%%%%%%% end figure %%%%%%%%%%%%%%%%%%%%%%%%%%%%%%%%%%%%%%%

If one adjusts the five parameters $G$, $\eta$, $J_e^0$, $\beta$ and $\tau_c$ to the shear data, they exhibit a strong scatter; the data are too featureless to be able to fix five parameters to the desired accuracy of about 10 percent. One way to overcome this difficulty is to fit both the shear and the dielectric data with the same $\tau_c$. This way was employed in our first two examples, PPE and DC704 (another diffusion pump oil). Then one obtains the other four parameters with reasonable accuracy from the shear data and $\epsilon(0)$, $\epsilon_\infty$ and $\tau_D$ from the dielectric data. One gets an excellent fit, as one can see in Fig. 3. The parameters $\beta$ and $f_0=GJ_e^0-1$ obtained turn out to be temperature-independent within their error bars. 

\begin{table}[h]
	\centering
		\begin{tabular}{|c|c|c|c|c|c|}
  \hline
  glass former          & PPE      & DC704   & TPE     & glycerol & TPG   \\  \hline
$T$ (K)                 & 254      & 217     & 250     & 192.5    & 192   \\  
$G$ (GPa)               & 1.067    & 1.602   & 1.19    & 3.49     & 2.40  \\  
$\eta$ (GPa s)          & 0.0634   & 0.216   & 0.210   & 9.44     & 3.61  \\  
$\tau_c/\tau_m$         & 2.37     & 2.01    & 2.2*    & 2.2*     & 7.0   \\  
$\tau_D/\tau_m$         & 8.1      & 8.8     & 5.9     & 7.5      & 19.0  \\  
$f_0$                   & 2.38     & 2.46    & 2.83    & 4.61     & 5.5   \\  
$\beta_{shear}$         & 0.64     & 0.61    & 0.53    & 0.66     & 0.64  \\
$\beta_\epsilon$        & 0.64     & 0.61    & 0.69    & 0.667    & 0.64  \\  \hline
		\end{tabular}
	\caption{Eshelby model parameters for five highly viscous liquids. The asterisk denotes set values.}
	\label{tab:tab1}
\end{table}

As it turns out, the value $f_0$ for Santovac 5P in Table I is a factor of 1.5 larger than the expectation of $\pi/2$ of the Eshelby model. The deviation is clearly out of the error bar of about 10 \%. This is in fact in agreement with shear compliance data in other molecular glass formers \cite{plazek}, which show a similar tendency to higher values. Obviously, the relaxational part of the recoverable compliance tends to be larger than the Eshelby expectation. We come back to this point in the discussion.

The measurements of the Roskilde group \cite{niss} comprise two other type-A glass formers (glass formers without a 
discernible Johari-Goldstein peak), TPE and DC704, which give similar results. However, in TPE it turned out to be 
necessary to allow for a sizable difference of the $\beta$ in dielectrics and shear, setting the ratio $\tau_c/\tau_m$ 
to the Eshelby expectation of 2.2 (see Table I).

Glycerol turned again out to be an example where shear and dielectrics could be fitted with the same $\beta$. Fig. 4 
shows the fit of the shear data, \cite{donthg}. Their $\beta$ agreed within the error bars with the one adjusted to 
dielectric data \cite{loidl,rossler}. Also, the times $\tau_c$ showed reasonable agreement, taking into 
account that these measurements come from different samples in different laboratories.

As soon as one has a pronounced Johari-Goldstein peak, the relation $\tau_c=2.2\tau_m$ does not seem to hold any longer. 
In tripropylene glycol (Fig. 5), even the fit of the shear data alone already requires a considerably higher $\tau_c$. 
This becomes even worse if one demands the same $\tau_c$ in dielectrics and shear. The fit in Fig. 5 was done with the 
same $\beta$ for dielectrics and shear, but with $\tau_c=6.9\tau_m$.

\section{Discussion and conclusions}

The comparison to experiment in the preceding section has raised more questions than it has answered. Insofar, the 
Eshelby model presented here is at present not more than one of the many alternatives in the field. It has, however, one 
distinct advantage, namely a direct connection to the elementary flow process.

To illustrate this, let us consider the example of Santovac 5P or polyphenylene of Fig. 3. The molecule consists of five 
phenyl rings bonded by oxygen atoms. Such an oxygen bond is rather flexible. Therefore one should assess the typical 
molecular volume to one of these phenyl rings, as far as packing considerations are concerned. The density is about 1250 
kg/m$^3$, so the volume of one phenyl ring is 0.12 nm$^3$.

With this volume and the modulus of 1.06 GPa, the formation energy $E_a$ of the four-molecule strain defect of Fig. 1 
according to eq. (\ref{ea}) is 0.173 eV, corresponding to a temperature of 2004 K. The barrier height should be substantially lower because of the anharmonicity of the potential. This barrier height is too low; the barriers which cause the flow are centered around $k_BT_g\ln{\tau_m/\tau_0}$, where $\tau_0$ is the 
microscopic time of about $10^{-13}$ s, which means a barrier height of 0.7 eV. On the other hand, the formation energy 
of the four-atom defect is high. In fact, it is too high; eq. (\ref{single}) tells us that to get a relaxational 
compliance of $\pi/2G$ (that is what we need according to eq. (\ref{j0e}) requires a number of symmetric-potential 
four-molecule defects $n_0$ per molecule given by
\begin{equation}\label{n0}
\frac{\pi}{2G}=\frac{n_0}{v_{mol}}\frac{v^2e_i^2}{16k_BT}.
\end{equation}
Here $v$ is four times the molecular volume $v_{mol}$ and $e_i$ is $\pi/3$. Thus we find that we need about 0.05 
four-atom defects per molecule to get the sample to flow. However, with a formation energy eight times higher than the 
glass temperature $T_g$, it is more likely we will get less than $10^{-3}$ such defects per molecule. In fact, the four-atom 
defect of Fig. 1 seems a better candidate for the tunneling states which determine the low-temperature properties than 
for the flow defects; it even has the right order of magnitude of coupling to the strain.

Looking at eq. (\ref{n0}), one realizes that the way out of this dilemma is to increase the volume of the core region. 
Let us consider a core region of hundred phenyl rings with a diameter of 2.8 nm with the same formation energy as the 
four-atom defect. Then its spontaneous shear deformation $e_i$ is only one fifth of $\pi/3$, a shear angle of 12 
degrees, but its coupling constant to the compliance is a factor of 25 higher. Therefore one needs a factor of 25 less 
defects to make the sample flow, approaching the numbers which one expects on the basis of their Boltzmann factor.

A second important point is the high barrier. This cannot be a defect which reaches its saddle-point by only a 
deformation by $e_i/2$; there must be a more complicated rearrangement requiring an energy about four times higher. If 
the saddle point is indeed only reached by a high-energy rearrangement of the core region, one begins to understand why 
the relaxational compliance is higher than the prediction of eq. (\ref{j0e}). There might be components of the motion 
which reverse after a while in following rearrangements, making the contribution of the defect to the flow smaller than 
expected. Imagine, for instance, that part of the rearrangement is a reorientation of an interstitial \cite{granato}. 
This would not contribute to the flow, but it would enhance the relaxational compliance. Another possibility is a 
molecular configuration change, which would also contribute to the compliance, but not to the flow, because it reverses 
after a while. The Johari-Goldstein peak is probably of this nature. But why the introduction of such a peak leads to a 
violation of the Eshelby lifetime condition, eq. (\ref{tauc}), as shown in the TPG example of Fig. 5, is not immediately clear.

Another unsolved riddle (an old riddle \cite{boettcher}) is the short Debye relaxation time. The ratio of $\tau_D$ and 
$\tau_m$ according to eq. (\ref{taud}) is
\begin{equation}
\frac{\tau_D}{\tau_m}=\frac{4\pi Gr_H^3}{k_BT}.
\end{equation}
If we take $r_H$ from the molecular volume by $4\pi r_H^3/3=v_{mol}$, this ratio for the phenyl ring in Santovac 5P at 
254 K should be 110, a factor of 14 higher than the one found in experiment. In terms of the radius $r_H$, this means 
that the radius must be only 40 percent of the one calculated from the molecular volume. One knows from NMR gradient 
measurements of the molecular diffusion \cite{chang,qi} that the hydrodynamic radius is a bit smaller than the one 
calculated from the molecular volume, but not that much. Maybe the process which we characterize by $\tau_D$ has another 
and faster mechanism than the one considered by Debye, which might account for the fact that it is accompanied by the 
relaxation of the energy and by the relaxation of the structure \cite{bow,bzow}.

A third unanswered question is whether the structural rearrangements fully relax the molecular orientation, as assumed 
for simplicity in the derivation of equs. (\ref{ep}) and (\ref{epp}). Their number is limited by the cutoff at $V_c$, so 
it is conceivable that there remains a rest of dielectric polarization, which should appear as an additional Debye 
relaxation at the Debye relaxation time. In fact, dielectric measurements in the monoalcohols \cite{richert} look as if 
most of the reorientation happened in one big Debye-like process long after the Maxwell time.

We conclude that the Eshelby model for the highly viscous flow, though it raises more questions than it answers, 
provides a new way to tackle an old and difficult problem on a quantitative level.

\end{document}